\documentstyle [11pt] {article}
\title {\bf COSMOLOGICAL MODELS WITH NONLINEARITY OF
       SCALAR FIELD INDUCED BY YANG-MILLS FIELD}

\author {V.K.Shchigolev\thanks{e-mail: shchigol@sv.uven.ru},~ M.V.Shchigolev}
\date {}

\def\ds {\displaystyle}
\def\sh {\mathop {\rm sinh}}
\def\ch {\mathop {\rm cosh}}

\begin {document}
\maketitle
\begin {center} Ulyanovsk State University, Institute for Theoretical
Physics \\
Ulyanovsks 432700, RUSSIA \\
\end {center}
\bigskip
\begin {center}
 {\small The exact solutions of  Einstein - Yang - Mills and interacting with SO (3) - Yang-Mills field nonlinear scalar field equations in a class of
spatially  homogeneous  cosmological Friedmann models  are obtained.} \\
\end {center}
\bigskip

PACS-94: 04.20. Jb \ 04.40.-b \bigskip

\section{Introduction}

It is known that one can construct the cosmological models of inflation basing
oneself on the solutions of the system of Einstein equations and nonlinear
(in the simplest variant , scalar) fields. Nonlinearity of the scalar field
is expressed in his self-acting through nonquadratic potential, which can
take different modifications in hierarchy of the inflationary cosmological
models \cite{C1}.
Nevertheless, there are no many examples of the exact solutions of Einstein
equations and nonlinear scalar field. A method of fine tuning \cite{C2}, which
allows in some cases to restore a dependence of the potential on the
field and in all cases to analyze the dependence of the field and the
potential on the time, is rather effective for this purpose. In cosmology
using nonlinearity of the scalar field induced with the interacting with the
other fields seems to be more realistic approach for describing a creation
of nonlinearity in the frame of the classical field theory. Unfortunately,
thus one has to solve difficult nonlinear system of the equation not only for scalar field, but for the field inducing
nonlinearity of this field. It is not always possible to receive exact
solutions even for the simplest field configurations. The situation becomes
more difficult, when one considers self-gravitating fields, meaning
a cosmological or astrophysical applications of the solutions.
Interesting solutions might be obtained when the cosmological gravitational
field has background character, as it was done, for example, in \cite{C3} and \cite{C4}.
But these solutions don't have direct relation to the problem of cosmological
scenario, even by virtue of their static character. Nevertheless, the idea
of induced nonlinearity is fairly attractive for realization, generally in
cosmology and probably in cosmology of early Universe especially. Earlier
in cosmology there were obtained some exact solutions for self-gravitating
Yang-Mills (YM) fields (look, for example, \cite{S1}). However if we use the obtained solutions
in this work to induce nonlinearity of scalar field in cosmology
it would lead to inhomogeneous configurations of the scalar field and
accordingly to inhomogeneous cosmological models. At the same time
nontrivial topology of SO(3)-YM fields gives us an extra possibilities
for searching the exact solutions of Einstein-YM and nonlinear scalar
field equations. In the present paper such system is solved in the frame of Friedmann
cosmological models and general spherically symmetric Wu-Yang ansatz for YM
fields \cite{C7}.

\section{Fundamental equations}

\rm Let us start from the Lagrangian of self-gravitating scalar field,
nonlinearity of which is induced by SO(3) YM field \cite{C3}:
$$
L=\frac{R}{2\symbol{26}}+\frac{1}{2}\varphi_{,\alpha}\varphi^{,\alpha}-
\frac{1}{16\pi}F_{\alpha\beta}^{a}F^{a\alpha\beta}\Psi(\varphi),\eqno (1)
$$
where $F_{ij}^{a}=\partial_{i}W_{j}^{a}-\partial_{j}W_{i}^{a}+
e\varepsilon_{abc}W_{i}^{b}W_{j}^{c}$ is tensor of YM field , and
$\Psi(\varphi)$ is the function of interaction of the
scalar field and YM field. Lets notice that in \cite{C4} the similar Lagrangian
is used for describing the interaction with the electromagnetic field and
for analyzing static configuration on a  background of static
Friedmann-Robertson-Walker  model. In this work the similar Lagrangian
is substantiated by the existence of the process  $\pi \to 2\gamma$ and
in paper \cite{S1} it is marked that in the case when  $\ds\Psi(\varphi)=
e^{-2\lambda\varphi}$ the Lagrangian Eq.(1) contains
Kaluza-Klein theory after compactification with $\ds\lambda=\sqrt{3}$ and the
superstrings theory for
$\ds\lambda=1$. The same type of interaction emerges in Brans-Dicke theory
with $\ds\lambda=1/\sqrt{2\omega+3}$.

  The equations for YM fields, the scalar field and the gravitational
potentials can be obtained by variation of the Lagrangian by the fields
and space-time metric. As a result we obtain the system of the proper
Einstein equations
$$
G_{\mu}^{\nu}=\symbol{26}T_{\mu}^{\nu},\eqno (2)
$$
where energy-momentum tensor has the following form
$$
T_{\mu}^{\nu}=\varphi_{,\mu}\varphi^{,\nu}-\frac{1}{4\pi}F_{\mu\beta}^a
F^{a\nu\beta}\Psi(\varphi)-\delta_{\mu}^{\nu}\left[ \frac{1}{2}
\varphi_{,\alpha}\varphi^{,\alpha}-\frac{1}{16\pi}F_{\alpha\beta}^{a}
F^{a\alpha\beta}\Psi(\varphi)\right],\eqno (3)
$$
YM equations:
$$
D_{\nu}\left(\sqrt{-g}F^{a\nu\mu}\Psi(\varphi)\right)=0\eqno (4)
$$
and the equations for the scalar field:
$$
\frac{1}{\sqrt{-g}}\frac{\partial}{\partial x^{\nu}}\left(\sqrt{-g}
g^{\nu\mu}\frac{\partial \varphi}{\partial x^{\mu}}\right)+
\frac{1}{16\pi}F_{\alpha\beta}^{a}
F^{a\alpha\beta}\Psi_{\varphi}=0.\eqno (5)
$$
Symbol $D_{\nu}$ means covariant derivative.

Let us assume that spacetime interval has spherically symmetry
properties and write it down in the form:
$$
 ds^2=dt^2-U(r,t)dr^2-V(r,t)d\Omega^2, \eqno (6)
$$
where
$$
 d\Omega^2=d\theta^2+\sin^2\theta d\phi^2. 
$$
Lets also write down Wu-Yang ansatz for YM fields
$$
W^a_0=x^a\frac{W(r,t)}{er},~W_i^a=\varepsilon_{iab}x^b\frac{K(r,t)-1}{er^2}+\left(\delta_{i}^a
-\frac{x^a{x_i}}{r^2}\right)\frac{S(r,t)}{er},\eqno (7)
$$
where isotopic indexes $a,b,...=1,2,3$ and space-time indexes
$i,j,k...=1,2,3$. $K,S$ and $W$ are the field functions of $r$ and $t$.
With help of orthonormalized isoframe
$$
\begin{array}{l}
{\bf n}=(\sin\theta\cos\varphi,\sin\theta\sin\varphi,\cos\theta),\\
{\bf l}=(\cos\theta\cos\varphi,\cos\theta\sin\varphi,-\sin\theta),\\
{\bf m}=(-\sin\varphi,\cos\varphi,0)
\end{array}
$$
in spherical system of coordinates ansatz Eq.(7) take the following form
$$
\begin{array}{l}
\ds{\bf W}_1=0\\
\ds{\bf W}_2=e^{-1}\left\{(K-1){\bf m}+S{\bf l}\right\},\\
\ds{\bf W}_3=e^{-1}\left\{-(K-1){\bf l}+S{\bf m}\right\}\sin\theta,\\
\ds{\bf W}_0=e^{-1}W{\bf n}.
\end{array} \eqno (8)
$$
As a result we have:
$$
\begin{array}{l}
\ds {\bf F}_{01}=-{\bf F}_{10}=-e^{-1}W'{\bf n},\\
\ds{\bf F}_{02}=-{\bf F}_{20}=
e^{-1}\left((\dot{K}+WS){\bf m}+(\dot{S}-WK){\bf l}\right),\\
\ds{\bf F}_{03}=-{\bf F}_{30}=
e^{-1}\left((\dot{S}-WK){\bf m}-(\dot{K}+WS){\bf l}\right)\sin\theta,\\
\ds{\bf F}_{12}=-{\bf F}_{21}=
e^{-1}\left(K'{\bf m}+S'{\bf l}\right),\\
\ds{\bf F}_{23}=-{\bf F}_{32}=
e^{-1}\sin\theta\left(K^2-1+S^2\right){\bf n},\\
\ds{\bf F}_{13}=-{\bf F}_{31}=
e^{-1}\sin\theta\left(-K'{\bf l}+S'{\bf m}\right).
\end{array} \eqno (9) 
$$
Here and below an overdot denotes the derivative
$\ds\frac{\partial}{\partial{t}}$
and the prime denotes the derivative $\ds\frac{\partial}{\partial{r}}$.
Taking into account ansatz for YM fields Eq.(9) Einstein equations Eq.(2)
for energy-momentum tensor Eq.(3) may be written down in the form
$$
\begin{array}{rcl}
\ds G_0^0&=&\ds \kappa_0 \left\{ \frac{\dot\varphi^2}{2}+\frac{\varphi'^2}{2}+
\frac{1}{8\pi e^2}\left[\frac{W'^2}{U}+\frac{2[(\dot K+WS)^2
+(\dot S-WK)^2]}{V}+\frac{2[K'^2+S'^2]}{UV}+ \right.\right.\\[15pt]

 & & \ds \left.\left. \frac{(K^2-1+S^2)^2}{V^2}
\right]\Psi\right\};  \\ [15pt]

\ds G_1^1&=&\ds \kappa_0 \left\{ -\frac{\dot\varphi^2}{2}-\frac{\varphi'^2}{2}+
\frac{1}{8\pi e^2}\left[\frac{W'^2}{U}-\frac{2[(\dot K+WS)^2+(\dot S-WK)^2]}{V}-
\frac{2[K'^2+S'^2]}{UV}+   \right.\right. \\ [15pt]

 & &\ds \left.\left. \frac{(K^2-1+S^2)^2}{V^2} \right]\Psi\right\};  \\ [15pt]

\ds G_2^2&=&\ds G_3^3=\kappa_0 \left\{ -\frac{\dot\varphi^2}{2}+\frac{\varphi'^2}{2}-
\frac{1}{8\pi e^2}\left[\frac{W'^2}{U}+\frac{(K^2-1+S^2)^2}{V^2}
\right]\Psi\right\};  \\ [15pt]

\ds G_1^0&=&\ds \kappa_0 \left\{ \dot\varphi\varphi'+
\frac{1}{2\pi e^2}\frac{K'(\dot K+WS)+S'(\dot S-WK)}{V}
\Psi\right\},
\end{array}\eqno (10)
$$
where the components of Einstein tensor $G_{\mu}^{\nu}$ are
calculated in metric Eq.(6).
The equations for YM fields Eq.(4) may be written down with help of
ansatz Eq.(9) in the following way
$$
\begin{array}{l}
\ds \frac{\partial}{\partial r}\left(\frac{VW'}{2\sqrt U}\Psi\right)
+\sqrt U\left(\left(\dot S-WK \right)K-\left(\dot K+WS \right)S\right)\Psi=0; \\
\\
\ds \frac{\partial}{\partial t}\left(\sqrt U\left(\dot S-WK\right)\Psi\right)-
\frac{\partial}{\partial r}\left(\frac{S'}{\sqrt U}\Psi\right)
+\sqrt U\left(\frac{\left(K^2-1+S^2 \right)S}{V}
-\left(\dot K+WS \right)W\right)\Psi=0; \\
\\
\ds \frac{\partial}{\partial t}\left(\sqrt U\left(\dot K+WS\right)\Psi\right)-
\frac{\partial}{\partial r}\left(\frac{K'}{\sqrt U}\Psi\right)
+\sqrt U\left(\frac{\left(K^2-1+S^2 \right)K}{V}
+\left(\dot S-WK \right)W\right)\Psi=0, \\
\end{array} \eqno (11)
$$
and the equation for the scalar field is
$$
\begin{array}{c}
\ds
\frac{1}{V\sqrt U}\frac{\partial}{\partial t}\left(V\sqrt U\dot\varphi\right)-
\frac{1}{V\sqrt U}\frac{\partial}{\partial r}\left(V \frac{\varphi'}
{\sqrt U}\right)-\\
\\
\ds -\frac{1}{8\pi e^2}\left[\frac{W'^2}{U}+\frac{2[(\dot K+WS)^2
+(\dot S-WK)^2]}{V}-\frac{2[K'^2+S'^2]}{UV}-\frac{(K^2-1+S^2)^2}{V^2}
\right]\Psi_{\varphi}=0,
\end{array} \eqno (12)
$$
where $\ds \Psi_{\varphi}=\frac{d\Psi}{d\varphi}$.

Lets notice that without the scalar field the system of equations (10)-(11)
for pure self-gravitating YM field has been solved in \cite{S3} in the Tolman
metric with the simplifying assumptions of the functions:$K=S=0$, but $W\ne 0$,.
The last means that YM field has an electrical charge which causes an inhomogeneity
of the cosmological model. If one starts from the assumptions similar to the
ones in quoted work then it seems impossible to solve entire system of
Einstein-Yang-Mills equations and equation of nonlinear scalar field
Eqs.(10)-(17). In work \cite{C4} the authors found the static solutions of the similar
field equations on the background of static Universe. One may use this way
with reference to our situation. In present work we suggest a new
substitution for the YM field functions, that let us investigate the system
Eqs.(10)-(17).

\section{Models with Friedmann-Robertson-Walker metric}

Friedmann-Robertson-Walker (FRW) interval for homogeneous and
isotropic Universe may be written down in the form
$$
ds^2=dt^2-a^2(t)\left(dr^2+\xi^2(r)d\Omega^2\right) \eqno (13)
$$
where
$$
\xi (r)=
\left\{
\begin{array}{lr}
\sin r, & k=+1, \\
r, & k=0,\\
\sh r, & k=-1,
\end{array}
\right.
$$
and $k=0,\pm 1$ is a sign of the curvature of hypersurface $t=const$.
So we may write  $U=a^2(t)$ and $V=a^2(t)\xi^2 (r)$
for metric Eq.(6). According to Eq.(18) it is not difficult to find nonzero
components of Einstein tensor \cite{C5}
$$
\begin{array}{l}
\ds G_0^0=\frac{3}{a^2}\left(\dot a^2+k\right),\\
\ds G_1^1=G_2^2=G_3^3=\frac{1}{a^2}\left(\dot a^2+2a\dot a+k \right).
\end{array}
$$

If we demand that $W=0$, $K=K(r)$, $S=S(r)$,  and homogeneity of the
scalar field $\varphi=\varphi(t)$, then 
Eqs.(11) and (12) will take the form
$$
\begin{array}{l}
\ds S''-\frac{(K^2-1+S^2)S}{\xi^2}=0, \\
\ds K''-\frac{(K^2-1+S^2)K}{\xi^2}=0,\\
\ds \frac{1}{a^3}\frac{\partial}{\partial t}\left(a^3\dot\varphi\right)+
\frac{1}{8\pi e^2a^4}\left[\frac{2\left(K'^2+S'^2\right)}{\xi^2}+
\frac{\left(K^2-1+S^2\right)^2}{\xi^4}\right]\Psi_\varphi=0.
\end{array} \eqno (14)
$$
The first two equations structurally coincide. It expresses a symmetry
of the Lagrangian concerning the following transformations
$$
\begin{array}{l}
K=P(r)\cos{\alpha},~~~S=P(r)\sin{\alpha},
\end{array} \eqno (15)
$$
where $\alpha$ is an arbitrary constant. From Eq.(14) we may obtain the equation
for the function  $P(r)$:
$$
P''(r)-\frac{\left(P^2(r)-1\right)P(r)}{\xi^2}=0. \eqno (16)
$$
For $k=\pm 1$, i.e. for the case of the parabolic and the hyperbolic models of the
Universe solutions of Eq.(23) are the following functions
$$
P(r)=\xi'(r)=\left\{
\begin{array}{lr}
\cos r, &  k=+1, \\
\ch r, &  k=-1.  \\
\end{array}\right.
\eqno (17)
$$
With help of obtained solutions of YM equations the independent Einstein
equations are reduced to the system
$$
\begin{array}{l}
\ds \frac{3}{a^2}\left(\dot a^2+k\right)=\symbol{26}\frac{\dot\varphi^2}{2}+
\frac{3\symbol{26}}{8\pi e^2}\frac{1}{a^4}\Psi(\varphi), \\
\\
\ds \frac{1}{a^2}\left(\dot a^2+2a \ddot a +k\right)=
-\symbol{26}\frac{\dot\varphi^2}{2}-
\ds \frac{\symbol{26}}{8\pi e^2}\frac{1}{a^4}\Psi(\varphi).
\end{array} \eqno (18)
$$
At the same time the equation of the scalar field may be written down in form
$$
\frac{1}{a^3}\frac{\partial}{\partial t}\left(a^3\dot\varphi\right)+
\frac{3}{8\pi e^2}\frac{1}{a^4}\Psi_\varphi=0. \eqno (19)
$$
It is easy to verify that only two equations of Eqs.(18) and (19) are
independent. That is why we should define one of the three functions
$a(t)$, $\varphi(t)$ or $\Psi(\varphi)$
to solve the system. It is the most natural way to define the function of
interaction. But it is not a necessary condition and depends on a concrete
context of the problem.

Lets notice an interesting feature of Einstein
equations Eq.(18), starting from which one may write an effective energy
density $\epsilon(t)$ and the pressure $p(t)$ in the following form:
$$
\epsilon(t)=\frac{\dot\varphi^2}{2}+
3\frac{1}{8\pi e^2 a^4}\Psi(\varphi),~~~ p(t)=\frac{\dot\varphi^2}{2}+\frac{1}{8\pi e^2 a^4}\Psi(\varphi).
$$
Therefore $\epsilon-3p=-\dot\varphi^2\le 0$.
The equation of state becomes $\epsilon=p$ as the equality  $\Psi=0$ is valid.
So the effective pressure is defined by inequality
$\frac{1}{3}\epsilon\le p \le \epsilon$.
By selecting of the function of interaction one may obtain a required
equations of state from the disiered interval for solving the concrete
problem of the cosmological scenario. It seems to us rather useful to apply
the considered model to discuss the problems of quintessence in spirit of
work \cite{C8} on the basis of the possible cosmic magnetic field influence on
the expansion of the Universe.

We may use the method of fine tuning \cite{C6} to obtain the exact solutions of
system Eqs.(18) and (19). It means that we define the evolution of the scale
factor $a(t)$. From the equation Eq.(18) one may find
$$
\begin{array}{l}
\ds \dot\varphi^2=-\frac{6}{\symbol{26}}\left(\left(\frac{\dot a}{a}\right)^2+
\frac{\ddot a}{a}+\frac{k}{a^2}\right), \\
\\
\ds \Psi=\frac{8\pi e^2}{\symbol{26}}a^2\left(2\dot a^2+a\ddot a+2k \right).
\end{array} \eqno (20)
$$

Looking at these equations we may say that the solution for the real scalar
field  exists only if
$$
\left(\frac{\dot a}{a}\right)^2+
\frac{\ddot a}{a}+\frac{k}{a^2}<0.
\eqno (21)
$$
Obviously the last means that the accelerated expansion of the
Universe with $\ddot a >0$ is not always possible and in case when $k=+1$
in the frame of
the present model there is no such expansion. If we include a perfect
fluid and a cosmological constant to the model then inequality Eq.(21) and the
conclusions of the possibility of an accelerated regime essentially differ
from the ones presented here. It is going to be investigated in future.

Let us find for example the solutions for nonlinear scalar field Eq.(20) using
the mentioned method. Let the scale factor $a(t)=a_0t$, and a sign
of the curvature $k=-1$.
Then from the equation Eq.(20) we obtain
$$
\begin{array}{l}
\ds \dot\varphi=\pm\sqrt{\frac{6}{\symbol{26}}\left(1-a_0^2\right)}\frac{1}{a_0t}, \\
\\
\ds \Psi=\frac{16\pi e^2}{\symbol{26}}a_0^2\left(a_0^2-1\right)t^2.
\end{array}
$$
From the first equation it follows that the solution exists if $a_0<1$
and may be written down in the form
$$
\varphi=\pm\sqrt{\frac{6}{\symbol{26}}\left(\frac{1}{a_0^2}-1\right)}\cdot \ln t
+\varphi_0.
$$
Excluding the time from the two last equations, we find the function of
interaction
$$
\Psi=\Psi_0e^{\pm 2\lambda\varphi}, \eqno (22)
$$
where $\ds \Psi_0=\frac{16\pi e^2}{\symbol{26}}a_0^2\left(a_0^2-1\right)
e^{\mp 2\lambda\varphi_0}$, and
$$
\ds \lambda=\sqrt{\frac{\symbol{26}}{6}
\frac{a_0^2}{(1-a_0^2)}}.
$$

Lets consider other example. Let us assume the harmonic dependence for
the evolution of the scale factor in the open model of the Universe:
$$
a(t)=H_0^{-1}\sin(H_0t).
$$
Getting this to the system Eq.(20) one may integrate this system and obtain
the following expression
$$
\begin{array}{l}
\ds\varphi=\pm\sqrt{\frac{12}{\symbol{26}}}H_0t+\varphi_0, \\
\\
\ds \Psi=-\frac{24\pi e^2}{\symbol{26}H_0^2 }\sin^4H_0t.
\end{array}
$$
With help of these formulas one may obtain the function of interaction in
the obvious form
$$
\ds \Psi=\Psi_0\sin^4\left[ \sqrt{\frac{\symbol{26}}{12}}(\varphi-\varphi_0)\right],\eqno (23)
$$
where $\ds \Psi_0=-\frac{24\pi e^2}{\symbol{26}H_0^2}$.

Lets notice that despite the concrete form of the function of interaction
and consequently the solutions of Eq.(18) according to formulas Eqs.(9),(15)
and (17) the nonzero components of the stress tensor take the
form
$$
\begin{array}{l}
\ds {\bf F}_{21}=-{\bf F}_{12}=ke^{-1}\xi (r)(\cos \alpha {\bf m} +
\sin \alpha {\bf l}),\\
\\
{\bf F}_{31}=-{\bf F}_{13}=ke^{-1}\xi (r)\sin \theta(\sin \alpha {\bf m} - \cos \alpha {\bf l}),\\
\\
{\bf F}_{32}=-{\bf F}_{23}=ke^{-1}\xi^2 (r)\sin \theta {\bf n}.
\end{array} \eqno (24)
$$
From Eq.(20) it follows that YM field has only magnetic components. It is
interesting that exactly this circumstance allowed us to find homogeneous
solutions for the scalar field. Really one may easily find from Eq.(24) the
invariant for
YM field $\ds {\bf F}_{\mu\nu}{\bf F}^{\mu\nu}=3e^{-2}a^{-4}(t)$, that
caused the dependence of nonlinear on a field
second term in the equation of the scalar field Eq.(18) only on time.
\section{Conclusion}

So, it has been shown that the system of Einstein-Yang-Mills equations and nonlinear
scalar field that is obtained from the Lagrangian Eq.(1) has the solutions with
homogeneous scalar field, which interacts with YM field Eq.(23) in Friedmann
Universe, limited only by one condition Eq.(21). The system reduced to two
independent Eqs. (18) or (20) and demands to be completed by the
concrete function of interaction or by definition of necessary rate of
the evolution of the scale factor. It was shown that realization of the
last one always may be carried out. And the simple examples Eqs.(22) and (23)
illustrate the possibility of revealing of the obvious dependence of the
function of interaction from the scalar field at some regime of Universe
expansion.
\\

{This work was supported by  Russian Basic Research Foundation via grants
No 98-02-18040 and  No 00-01-00260.}

\end{document}